# Topological Phase Transition-Induced Tri-Axial Vector Magnetoresistance in $(Bi_{1-x}In_x)_2Se_3$ Nanodevices

*Minhao Zhang,*[†,∇] *Huaiqiang Wang,*[‡,∇] *Kejun Mu,*[§] *Pengdong Wang,*[§] *Wei Niu,*[†] *Shuai Zhang,*[‡] *Guiling Xiao,*[⊥] *Yequan Chen,*[†] *Tong Tong,*[†] *Dongzhi Fu,*[‡] *Xuefeng Wang,**[†,#] *Haijun Zhang,**[‡,#] *Fengqi Song,**[‡,#] *Feng Miao,*[‡,#] *Zhe Sun,*[§] *Zhengcai Xia,*[⊥] *Xinran Wang,*[†,#] *Yongbing Xu,*[†,#] *Baigeng Wang,*[‡,#] *Dingyu Xing*[‡,#] and *Rong Zhang*[†,#]

[†]National Laboratory of Solid State Microstructures and Jiangsu Provincial Key Laboratory of Advanced Photonic and Electronic Materials, School of Electronic Science and Engineering, Nanjing University, Nanjing 210093, China

[‡]National Laboratory of Solid State Microstructures, School of Physics, Nanjing University, Nanjing 210093, China

[§]National Synchrotron Radiation Laboratory, University of Science and Technology of China, Hefei 230029, China

[⊥]Wuhan National High Magnetic Field Center, Huazhong University of Science and Technology, Wuhan 430074, China

[#]Collaborative Innovation Center of Advanced Microstructures, Nanjing University, Nanjing 210093, China

*Corresponding authors. E-mail: xfwang@nju.edu.cn; zhanghj@nju.edu.cn; songfengqi@nju.edu.cn.

**ABSTRACT:** We report the study of a tri-axial vector magnetoresistance (MR) in nonmagnetic $(Bi_{1-x}In_x)_2Se_3$ nanodevices at the composition of $x = 0.08$. We show a dumbbell-shaped in-plane negative MR up to room temperature as well as a large out-of-plane positive MR. MR at three directions is about in a -3%: -1%: 225% ratio at 2 K. Through both the thickness and composition-dependent magnetotransport measurements, we show that the in-plane negative MR is due to the topological phase transition enhanced intersurface coupling near the topological critical point. Our devices suggest the great potential for room-temperature spintronic applications, for example, vector magnetic sensors.

Topological phase transition (TPT)[1-6] in topological insulators (TIs)[7-9] is associated with changes in bulk topological invariants as well as topological surface states (SSs). The TPT in TIs has attracted a great deal of attention and caused a surge in efforts to research the exotic ground states, such as massive Dirac fermions[2] and topologically trivial SSs.[4] To date, extensive magnetotransport measurements have been performed on TIs and the intriguing physical phenomena have been identified, such as weak antilocalization,[10,11] Shubnikov-de Haas (SdH) oscillations,[12-14] universal conductance fluctuations,[15] quantum Hall effect[16,17] and the anisotropic magnetoresistance (MR).[18-22] Recently, as triggered by the discovery of topological semimetals,[23,24] the negative MR in nonmagnetic systems has attracted the renewed interest. An intriguing phenomenon is the negative longitudinal MR observed in TIs when a magnetic field is applied along the current direction.[25-28] However, TPT-induced tri-axial vector MR has never been reported.

In three-dimensional (3D) TIs, suitable atomic doping can effectively tune the spin-orbit coupling (SOC) strength and subsequently drive the bulk TPT.[1-6] For example, in $(Bi_{1-x}In_x)_2Se_3$ system,[3,5] a broad tunability from a $Z_2$ topologically insulator into a topologically trivial band insulator was revealed by the angle-resolved photoemission spectroscopy (ARPES). More recently, $(Bi_{1-x}In_x)_2Se_3$ thin films were shown to have two separate transitions using the temperature-dependent resistance measurements.[6]

Here, by intentional indium doping, we have driven the TPT in the nonmagnetic $(Bi_{1-x}In_x)_2Se_3$ series, as revealed by the ARPES spectroscopy. At composition of $x = $ 0.08 in the vicinity of the topological critical point (TCP), we observe a tri-axial vector MR in thin flakes up to room temperature, in which the MR depends on the relative orientation between the current and the magnetic field. We observe a large positive MR of ~225% in the out-of-plane magnetic field at 2 K. When in the in-plane magnetic field, we observe a negative MR (about -3% at $\varphi = 0°$ and -1% at $\varphi = 90°$). The negative MR shows to be very sensitive to the surface contribution, which is due to the spin-related effect induced by the TPT-enhanced intersurface coupling.

**RESULTS AND DISCUSSION**

(Bi$_{1-x}$In$_x$)$_2$Se$_3$ single crystals with nominal doping concentrations ($x$ = 0, 0.08, 0.10, 0.16 and 0.20) were synthesized by the melting method. The bulk single crystals were then exfoliated to form thin flakes (see Methods). Bi$_2$Se$_3$ has a simple surface band structure as opposed to an alloy such as Bi$_{1-x}$Sb$_x$,[29] while $\alpha$-In$_2$Se$_3$ is a topologically trivial compound with a direct band gap of 1.36 eV. The (Bi$_{1-x}$In$_x$)$_2$Se$_3$ series share a common rhombohedral structure with that of Bi$_2$Se$_3$, as revealed by X-ray diffraction and Raman spectroscopy (Figure S1a and b). Evidence for the occurrence of TPT is seen in the ARPES mapping around the $\Gamma$ point (Brillouin-zone center) in the (Bi$_{1-x}$In$_x$)$_2$Se$_3$ series (Figure 1). We scrutinize the width of the bulk gap and the surface band dispersion to see if the composition of $x$ = 0.08 is close to the TCP. Figure 1a shows that Bi$_2$Se$_3$ has a large inversion bulk gap (~0.3 eV), in good agreement with its theoretical value. With doping concentration of $x$ = 0.08, a notable shrinkage of the bulk gap (~0.1 eV) is observed (Figure 1b), which is likely to be a signature near the TCP. Far from the TCP at $x$ = 0.16, a subsequent increase in bulk gap is seen in Figure 1c.

The TPT occurs in our $(Bi_{1-x}In_x)_2Se_3$ system, as revealed by the closure of the bulk gap and its reopening near the TCP. Apart from the detailed evolution of the bulk gap, the TPT can also clearly be recognized by the surface band dispersion inside the bulk gap around the $\Gamma$ point. A gapless Dirac SS is explicitly seen across the bulk gap for $x = 0$ (Figure 1a), whereas it totally vanishes at $x = 0.16$ (Figure 1c), indicating that a topologically nontrivial phase transfers into a topologically trivial one as In doping content increases. We find that the SS still exists at $x = 0.08$ even though the system is very near the TCP. A blurry intensity can be observed inside the bulk gap at $x = 0.08$ by ARPES (Figure 1b), which is a direct manifestation of the SS. The position of Fermi level is deep inside the conduction band, as seen from the ARPES mapping. The charge polarity of bulk and SS is therefore $n$ type.

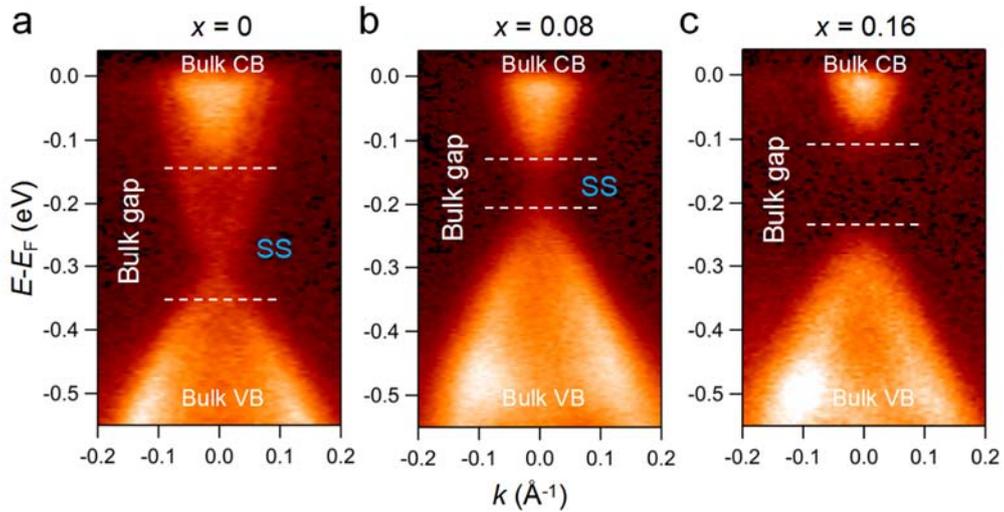

**Figure 1.** The bulk TPT in $(Bi_{1-x}In_x)_2Se_3$ system. (a to c) ARPES mapping of $(Bi_{1-x}In_x)_2Se_3$ series with $x = 0$, 0.08, and 0.16. CB and VB denote the conduction band and valence band, respectively.

Both in-plane and out-of-plane magnetotransport measurements were performed on thin flakes (with composition at $x = 0.08$ and thicknesses of ~40-90 nm) which were patterned by the six-terminal Hall-bar electrodes, as shown in Figure 2a (see Methods). The temperature-dependent resistance curve in Figure 2b shows a metallic transport behavior above 10 K followed by a residual resistance at low temperatures in Nanodevice 1. The resistivity of the sample at $x = 0.08$ is about 0.024 $\Omega\cdot$cm at room temperature. The metallic transport behavior corresponds to the position of Fermi level, as seen from the ARPES mapping, which is deep inside the conduction band. The transport evidence of SS for $x = 0.08$ is confirmed by angle-dependent SdH oscillations and the nontrivial Berry phase. In Figure 2c, we plot the sheet resistance ($R_{xx}\square$) and Hall resistance ($R_{xy}$) as function of the magnetic field $B$ up to 14 T. SdH oscillations are superimposed on the $R_{xx}\square(B)$, while $R_{xy}$ is found to be nonlinear, suggesting the presence of multiple channels. The nonlinear $R_{xy}$ has been shown in $(Bi_{1-x}In_x)_2Se_3$ system in the nontrivial phase with the coexistence of the bulk and SS.[3] While the $(Bi_{1-x}In_x)_2Se_3$ system is in the trivial phase with only one bulk channel, the $R_{xy}$ should appear linear.[3] We also observe the linear Hall resistance ($\theta = 0°$) in the topologically trivial phase at $x = 0.20$ at 2 K (Figure S2b), suggesting the disappearance of the surface channel as In doping content increases. After subtracting the smooth background, the SdH beats coincide with each other against perpendicular components of $B$ (Figure 2d). The electronic states are thus well confined in the plane of the thin flake, demonstrating the presence of the two-dimensional (2D) SS.[12,13,30] In Figure 2e, using the local maximum of $\Delta R$ to mark the integer Landau index ($n$), the intercept value of 0.528 is extracted from the Landau fan diagram, indicating the nontrivial $\pi$ Berry phase.[12,13,30,31]

Hence, we conclude that the composition of *x* = 0.08 is still in the topologically nontrivial phase, which hosts the SS.

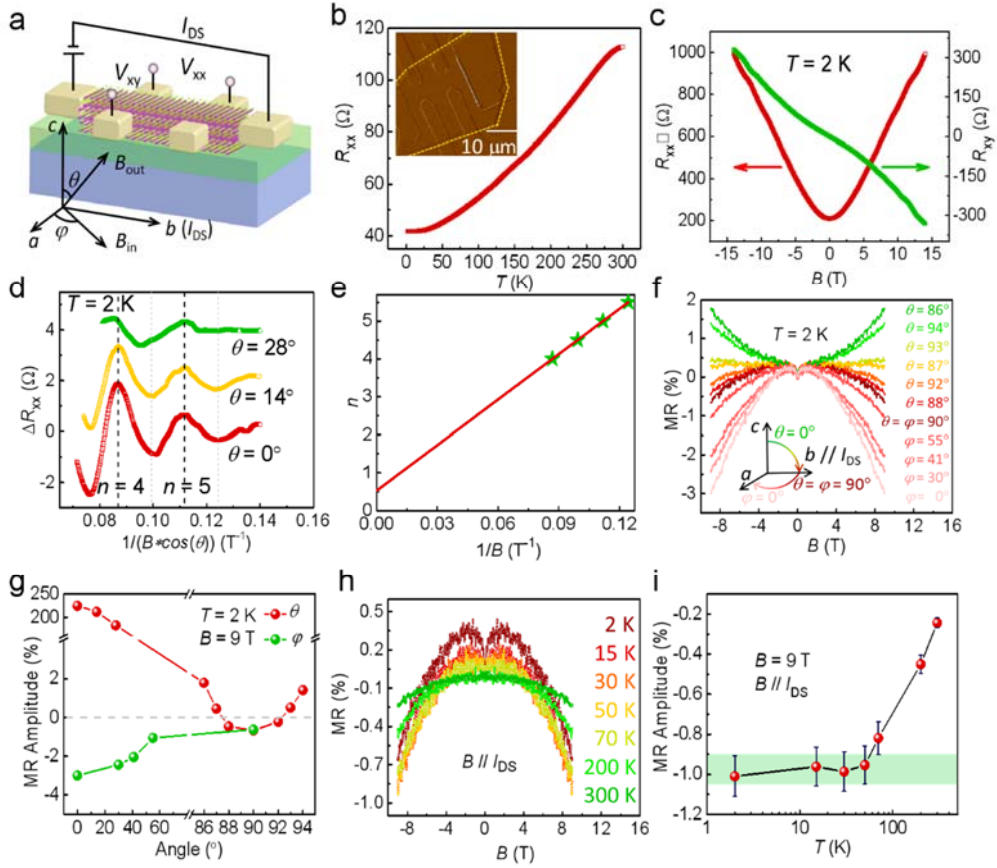

**Figure 2.** Room temperature vector MR at composition *x* = 0.08 in vicinity of the TCP at the topologically nontrivial side. (a) Schematic drawing of nanodevice structure and magnetotransport measurements. The current is applied on the two outside electrodes, where the inner two are for voltage detection. The out-of-plane angle between the magnetic field $B_{out}$ and coordinate axis *c* is defined as $\theta$. The in-plane angle between the magnetic field $B_{in}$ and coordinate axis *a* is defined as $\varphi$. (b) Temperature-dependent resistance curve. The inset contains an image of a nanodevice. The dashed lines are the sketch of the thin flake edges. (c) $R_{xx\square}$ and $R_{xy}$ as a function of *B* at 2 K. (d) SdH oscillations plotted against $1/(B\cos\theta)$ at the selected tilted angle $\theta$ after subtracting the smooth background at 2 K, indicating the 2D SS transport behavior. (e) The Landau fan diagram. The intercept is ~0.528, demonstrating the existence of the topologically nontrivial state. (f) Angle-dependent MR curve at 2 K. The negative MR signal is clearly observed when *B* // $I_{DS}$ at $\theta$ = 90°. It is much enhanced when the magnetic field is rotated away from $\varphi$ = 90°. (g) Graph of the angle ($\theta$ and $\varphi$) dependent MR amplitude at *B* = 9 T. For the angle $\theta$ varying from 0° to 94°, there is a tip seen in the red line which points to the negative MR very close to 90°. The negative MR amplitude reaches a maximum at about $\varphi$ = 0°, as seen from the green line. (h) Temperature-dependent negative MR when *B* //

$I_{DS}$. (i) Graph of the temperature-dependent negative MR amplitude. The negative MR is maintained up to 300 K.

In Figure 2f, we plot the MR as a function of magnetic field $B$ for different $B$ orientations in 3D space. The MR (%) is defined as $\frac{R(B)-R_0}{R_0} \times 100\%$, where $R_0$ is the resistance at zero field. The MR is found to have big differences for out-of-plane ($\theta$) and in-plane ($\varphi$) angles. At $\theta = 90°$ ($B \parallel I_{DS}$), we clearly observe a negative MR, where the MR begins to decrease at approximately ±1 T and continues over the entire magnetic field range (until ±9 T). The sharp cusp of MR within ±1 T originates from the weak antilocalization at low magnetic field. It could be induced by the strong SOC of the bulk.[24] The coupling of the SSs may also induce an additional correction to the conductivity under the low parallel field.[32] The negative MR exhibits the strong angular sensitivity at $\theta = 90°$ and a suppressed signal following a slight rotation of the magnetic field, as seen at $\theta = 87°$. The negative MR could even be enhanced further when $B$ is rotated in-plane from $\varphi = 90°$ ($B \parallel I_{DS}$) to $\varphi = 0°$ ($B \perp I_{DS}$). We show the negative MR in other multiple nanodevices, as summarized in Table 1. Its amplitude of ~-1% at 9 T is read in all five representative nanodevices with thicknesses of ~40-90 nm in the parallel magnetic field. The Nanodevice 3 is measured by 2-terminal geometry, which also shows the negative MR (Figure S3). The resistance of Nanodevice 3 is about 33 Ω at 2 K. While the resistance of Nanodevice 1 (4-terminal) is about 40 Ω at 2 K. Hence, the contact resistance in our devices is small. According to the ARPES mapping, the sample with $x = 0.08$ is heavily doped with the Fermi level in the conduction band (Figure 1b). Besides, the $R$-$T$ curve (Figure 2b) shows a metallic transport behavior.

Thus, the sample is just like a metal. We achieve the Ohmic contacts in our devices. Ohmic contacts are also achieved in 2-terminal TI devices at room temperature.[33] Further, our samples are fabricated with different geometry by the mechanical exfoliation. Therefore, the contact and sample geometry should be irrelevant to the in-plane anisotropy of the measurements. The spatial evolution of MR is more clearly illustrated by plotting the negative MR amplitude at $B = 9$ T at the different $\theta$ and $\varphi$ angles (Figure 2g). The MR amplitude (~225% at $\theta = 0°$) decreases considerably before reaching a minimum at $\theta = 90°$ as $\theta$ varies from 0° to 94°. The strongest negative MR signal (~-3%) is observed at $\varphi = 0°$ when $\varphi$ varies from 90° to 0°. We read the negative MR amplitudes at three directions of about -3% ($\varphi = 0°$), -1% ($\varphi = 90°$) and 225% ($\theta = 0°$) at 2 K. Such a TPT-induced tri-axial anisotropic MR has never been reported in nonmagnetic materials. Figure 2h shows the temperature dependence of the negative MR over a wide range from 2 to 300 K. The negative MR amplitude maintains up to 50 K and remains sizeable at room temperature, at ~-0.2% (Figure 2i), which provides the potential applications in vector magnetic sensors.

**Table 1.** Summary of the in-plane negative MR amplitude at 9 T ($B \parallel I_{DS}$) in five representative nanodevices.

| Nanodevice # | 1 | 2 | 3 | 4 | 5 |
|---|---|---|---|---|---|
| Thickness (nm) | 40 | 80 | 50 | 70 | 90 |
| Negative MR (%) | -0.62 ±0.031 | -0.51 ±0.026 | -0.73 ±0.037 | -1.24 ±0.062 | -0.99 ±0.050 |

In Nanodevice 2, the temperature-dependent resistance curve (Figure 3a) shows a metallic transport behavior. The negative MR shows a strong in-plane anisotropy as that of Nanodevice 1 (Figure 3b). The SdH oscillations are superimposed onto the negative MR in the high magnetic field, and obviously suppress the negative MR signal, as seen from the curvature of the smoothed solid line. It may come from the bulk carriers due to the angle-independent oscillations. The carriers tend to be localized in the cyclotron orbit with increasing magnetic field, which contributes to a positive MR component in the negative MR. In Figure 3c, the $\varphi$ angle-dependent negative MR amplitude at $B = 9$ T shows a period of 180°, demonstrating the strongest negative MR signal at $\varphi = 0°$ and 180° ($B \perp I_{DS}$). The polar diagram of the MR of Figure 3c shows the intriguing dumbbell-shaped patterns with two-fold symmetry from $\varphi = 0°$ to 360°, as shown in Figure 3d.

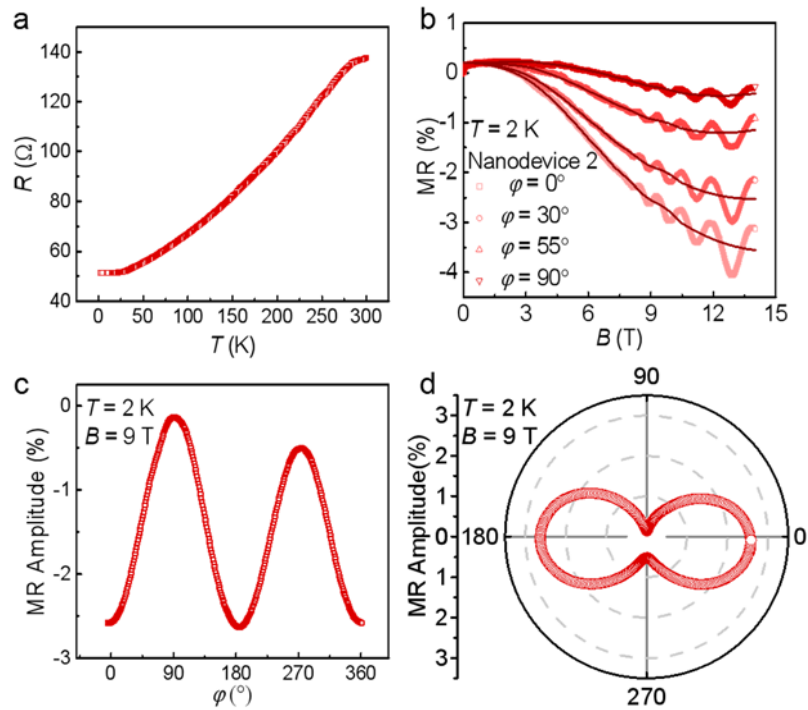

**Figure 3.** The dumbbell-shaped in-plane negative MR. (a) Temperature-dependent resistance curve of Nanodevice 2. (b) The $\varphi$ angle-dependent negative MR at 2 K in the Nanodevice 2. Obviously, the SdH oscillations suppress the negative MR signal, as seen from the curvature of the smoothed solid line at the high magnetic field. (c) The $\varphi$ angle-dependent negative MR amplitude at $B = 9$ T. (d) The polar plot of the negative MR amplitude at $B$ = 9 T of (b), illustrating the intriguing dumbbell-shaped negative MR.

We performed the controlled experiments on devices fabricated from thin flakes with argon plasma etching. We show that plasma exposure is a straightforward method of decreasing the thickness of our exfoliated samples. The thickness of pristine sample by mechanical exfoliation is more than ~40 nm. The exfoliated thin flakes are discretely distributed on the SiO$_2$/Si substrate. After the argon ion etching, the thickness of the etched sample is around 15 nm according to Figure S4b. The crystalline ordering of etched samples maintains as shown by the Raman spectra (Figure S4a). However, the surface becomes inhomogeneous with a roughness of ~1 nm (Figure S4b). The temperature-dependent resistance curve in Figure 4a also shows a metallic transport behavior in the etched sample. In Figure 4b, at $\theta$ = 0° ($B \perp I_{DS}$), the MR is found to have big differences between the etched and the pristine samples. The pristine sample exhibits a large nonsaturating MR, which is positive and becomes highly linear above 3 T, as shown by the dashed line. The linear MR vanishes altogether in bulk single crystals with a small surface-to-volume ratio (Figure S5). The linear MR is thus intrinsically linked to the 2D SS.[34] However, in the etched samples with argon plasma exposure, the linear MR and SdH oscillations are not observed (Figure 4b). By the logarithmic plot of Figure 4b, it can be clearly seen that the MR amplitude (~10%) of the etched sample is greatly suppressed compared with that (~375%) of the pristine sample (Figure 4c). The change of MR is due to the disorder and surface roughness generated by plasma etching, which may play an important role in the transport behavior.[35]

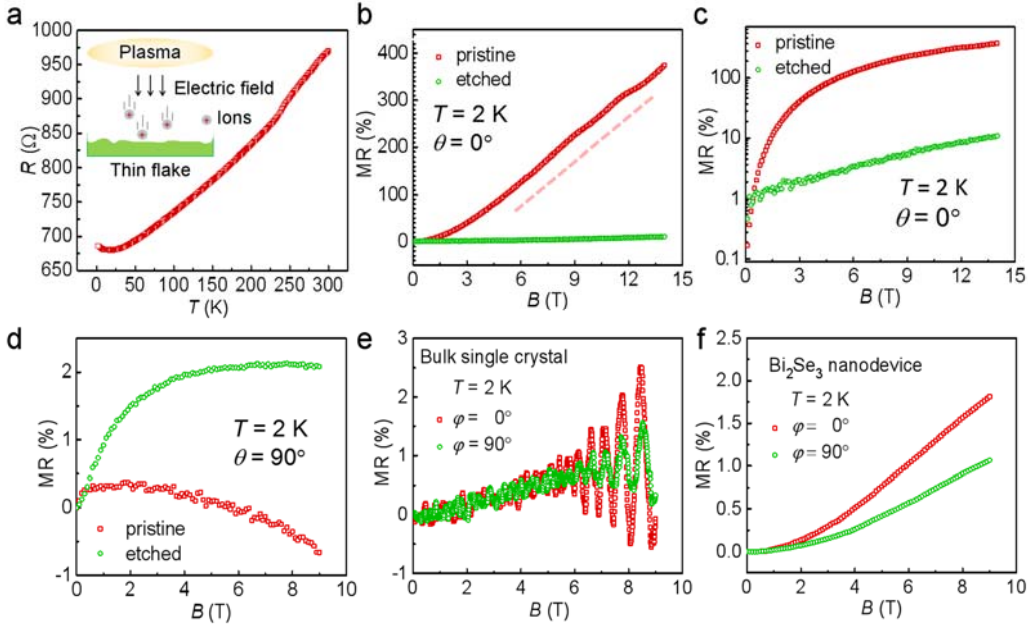

**Figure 4.** The surface-related negative MR near the TCP. (a) Temperature-dependent resistance curve. The inset contains a schematic diagram of the argon plasma etching. (b) The MR of the etched sample in contrast to the nonsaturating linear MR of the pristine sample in the out-of-plane magnetic field ($\theta = 0°$) at 2 K. (c) Logarithmic plot of MR of (b). The MR amplitude of the etched sample (~10%) is suppressed greatly compared with that of the pristine sample (~375%) at $B = 14$ T. (d) The positive MR of the etched sample in the in-plane magnetic field ($\theta = \varphi = 90°$) at 2 K. (e and f) The positive MR of the bulk single crystal and undoped $Bi_2Se_3$ thin flake (~60 nm) at 2 K, respectively, both in the in-plane magnetic field ($\varphi = 0°$ and 90°).

Figure 4d shows the MR at $\theta = 90°$ ($B // I_{DS}$) of the etched and the pristine samples, also displaying a distinct difference between them. The etched sample instead shows a positive MR in contrast to the observed negative MR in the pristine sample, which is again ascribed to the disorder and impurity after the argon plasma exposure. For a bulk single crystal in an in-plane magnetic field, the MR is displayed as positive either at $\varphi = 90°$ or 0°, superimposed with the predominant SdH oscillations (Figure 4e). Thus, the thickness provides the crucial role in the occurrence of the negative MR. Finally, we measured the magnetotransport properties of an undoped $Bi_2Se_3$ thin flake (~60 nm). The MR of $Bi_2Se_3$ just shows an in-plane positive MR either at $\varphi = 90°$ or $0°$ (Figure 4f). At $x = 0.20$ in the topologically trivial phase, we also observe an in-plane positive MR either at $\varphi = 90°$ or $0°$ (Figure S2, c and d), further corroborating the TPT-induced negative MR at $x = 0.08$ near the TCP.

As shown above, we find that the in-plane anisotropic negative MR takes place in exfoliated thin flakes with the pristine surface quality rather than the bulk near the TCP. Before understanding of the negative MR near the TCP, we rule out a few possible origins. Firstly, the hexagonal snowflake Fermi surface of the SSs in 3D TIs may induce a negative MR with six-fold symmetry.[36] This runs counter to our observation of the two-fold symmetry (Figure 3d). Secondly, the Fermi level is deep inside the conduction band (Figure 1b) which leads to a metallic transport behavior (Figure 2b). Hence, in contrast to compensated TIs of an insulating behavior, percolating puddles should not be involved in our observed negative MR.[26] Thirdly, no upturn at low temperatures below 10 K in the temperature-dependent resistance curve is observed (Figure 2b), inconsistent with weak localization effect that could be also responsible for the negative MR. Fourthly, in the case of the ultra-quantum limit, the ionic scattering could also induce the negative MR.[37] To rule out this origin, the Landau-level indices calculated from the high-pulsed-field measurements up to 60 T show that the electron transport remains in the semi-classical limit at $B \leq 14$ T (Figure S6). Finally, the recently notable Adler-Bell-Jackiw (ABJ) anomaly[24] observed in Weyl semimetals, which relies on the number imbalance of chiral fermions for $B // I_{DS}$, could also be the origin of the negative

MR. Instead, our observation of the strongest negative MR at $B \perp I_{DS}$ (Figure 2f) rules this out. For the ABJ anomaly, a nonzero $\boldsymbol{E}\cdot\boldsymbol{B}$ is a prerequisite for the chirality imbalance.[24]

Now we return to the origin of the anomalous in-plane negative MR. It could be interpreted by TPT-enhanced intersurface coupling and the underlying scattering mechanism related to the spin polarization of the SSs. The coupling of the SSs has been proven by the terahertz spectroscopy, which plays an important role in the change of conductance in $(Bi_{1-x}In_x)_2Se_3$ near the TCP.[38,39] The penetration depth of the SS wavefunction is proportional to $\hbar v_F/\Delta$,[40] where $\hbar$ is the reduced Planck constant, $v_F$ is the Fermi velocity, and $\Delta$ is the bulk gap. Given the small $\Delta$ (~0.1 eV) in $(Bi_{1-x}In_x)_2Se_3$ ($x = 0.08$) near the TCP (Figure 1b), it would be expected to have the lower threshold for the overlap of the wavefunctions of the top and bottom surfaces. Note that the thicknesses of our thin flakes range from 40 to 90 nm (Table 1); we thus infer that electrons in the top and bottom surfaces of the thin flakes can be strongly coupled (Figure 5a).

Once the SSs come to hybridize, the net spin polarization ($S_{net}$) would be generated proportional to the in-plane magnetic field ($\boldsymbol{B}_{in}$), as illustrated in Figure 5b. When the magnetic field is increased, the electron spins gradually align along the $\boldsymbol{B}_{in}$. Due to the aligned spins, the electron scattering is expected to be suppressed, and thus drives the negative MR.[41] The anisotropy is supposed to originate from the in-plane angle-dependent scattering probability during electron transport.[22] Moreover, the $S_{net}$ should be unrelated to the current density. This is supported by our variable current-density experiments, in which we observe the coincidence of the negative MR (Figure S7). In the out-of-plane magnetic field, the negative MR is overwhelmed by the positive MR, which is generally attributed to the deflection by the Lorentz force and localization by the cyclotron orbit. In the thin etched samples, the disorder and impurity make the interface coupling somewhat difficult. Thus, we observe a positive MR in the in-plane magnetic field (Figure 4d).

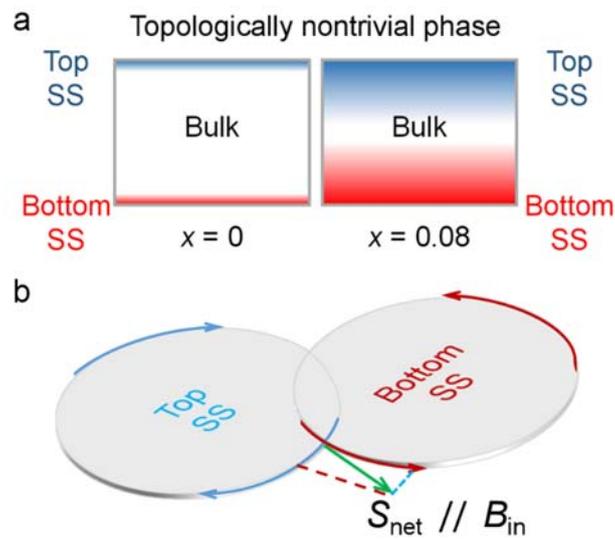

**Figure 5.** Schematic diagrams of spin polarization due to the TPT-enhanced intersurface coupling. (a) The depicted penetration depth of SS in thin flakes at $x = 0$ and 0.08 for comparison. The SS wavefunctions at $x = 0.08$ penetrate into the bulk much more deeply. (b) Net spin polarization ($S_{net}$) is generated proportional to the applied in-plane magnetic field ($B_{in}$). While increasing the field, the electron spins gradually align along the $B_{in}$. Thus, the electron-spin scattering is greatly suppressed.

**CONCLUSIONS**

We observe an in-plane angle-sensitive negative MR as well as an out-of-plane positive MR in $(Bi_{1-x}In_x)_2Se_3$ nanodevices, which reveals TPT-induced magnetotransport properties. Controlled experiments on etched samples and macroscopic single crystals demonstrate that the negative MR is associated with the TPT-enhanced intersurface coupling. Our experimental observations not only provide a route to produce the negative MR phenomenon through TPT, but also light the potential applications in various spintronic devices, particularly in vector magnetic sensors.

**METHODS**

**Materials growth and characterization.** $(Bi_{1-x}In_x)_2Se_3$ single crystals were grown by melting high-purity (99.99%) elements of Bi, In, and Se at 850 °C for one day in evacuated quartz tubes, which were then annealed at 850 °C for two days followed by cooling to room temperature for four days.[31] The crystalline structure of $(Bi_{1-x}In_x)_2Se_3$ series were determined by X-ray diffraction using a Cu K$\alpha$ line (Rigaku Ultima III) and a micro-Raman spectrometer (NT-MDT nanofinder-30) with a 514.5 nm Ar$^+$ laser, respectively. ARPES measurements on cleaved crystals were performed at National Synchrotron Radiation Laboratory in University of Science and Technology of China. We conducted the ARPES measurements on the cleaved (001) face of the bulk single crystals at 20 K to trace the evolution of electronic states across the TCP.

**Device fabrication.** The $(Bi_{1-x}In_x)_2Se_3$ thin flakes (with thicknesses ~40-90 nm) were exfoliated and transferred on doped Si substrates with 300 nm $SiO_2$, like the transfer process of graphene. The samples were subsequently patterned using standard Hall-bar electrodes with photolithography. Au electrodes (~50 nm) were deposited by magnetron sputtering. The thickness of the exfoliated samples was measured by an atomic force microscope (Asylum Cypher) operated in the tapping mode with a Si tip. The dry etching of the $(Bi_{1-x}In_x)_2Se_3$ thin flakes was carried out in a planar-type inductively coupled plasma system using Ar$^+$ plasma with a total gas flow rate of 30 sccm, a pressure of 1 Pa and a total etching time of 420 s. Then, surface morphology of the etched $(Bi_{1-x}In_x)_2Se_3$ thin flakes was examined by atomic force microscopy.

**Transport measurements.** Transport measurements were performed using a Quantum Design physical properties measurement system (PPMS) which can sweep the magnetic field between ±14 T at temperatures down to 2 K. The amplitude of the excitation current density was optimized to maintain the sufficient signal-to-noise ratio. The high-field measurements were performed in a pulsed magnetic field (up to 60 T) at Wuhan National High Magnetic Field Center, China. The MR presented in this work was obtained by symmetrization of the raw magnetotransport data with respect to the magnetic fields.

**ASSOCIATED CONTENT**

The authors declare no competing financial interest.

**Supporting Information**

Supporting Information Available: The in-plane positive MR of $(Bi_{1-x}In_x)_2Se_3$ thin flakes ($x = 0.20$) at 2 K; the in-plane positive MR of $(Bi_{1-x}In_x)_2Se_3$ thin flakes ($x = 0.20$) at 2 K; the negative MR in Nanodevice 3 measured by 2-terminal geometry; Raman spectra of $(Bi_{1-x}In_x)_2Se_3$ thin flakes ($x = 0.08$) after plasma exposure; nonsaturating parabolic MR of the bulk single crystal ($x = 0.08$) at 2 K; magnetotransport results of composition at $x = 0.08$ in the high-pulsed magnetic field; current-density dependence of the in-plane negative MR at 2 K. This material is available free of charge *via* the Internet at http://pubs.acs.org.

## AUTHOR INFORMATION

**Corresponding authors**

*E-mail: xfwang@nju.edu.cn.

*E-mail: zhanghj@nju.edu.cn.

*E-mail: songfengqi@nju.edu.cn.

**Author contributions**

∇M.Z. and H.W. contributed equally to this work. X.-f.W. conceived the study and designed the experiments. M.Z. prepared the samples, fabricated the devices and performed the electrical measurements. H.W. and H.Z. conducted the theoretical calculations. K.M. and P.W. performed the ARPES measurements. W.N., S.Z., Y.C., T.T., and D.F. performed the partial controlled experiments. G.X. carried out the pulsed-field magnetotransport measurements. F.S., F.M., Z.S., Z.X., X.-r.W, Y.X., B.W., D.X., and R.Z. contributed to the data analysis. X.-f.W. and M.Z. wrote the paper. All the authors discussed the results and commented on the manuscript.

## ACKNOWLEDGMENTS

We thank Prof. Chao-Xing Liu and Prof. Jinsheng Wen for some helpful discussions. This work was financially supported by the National Key Basic Research Program of China under grant no. 2014CB921103, 2017YFA0303203, and 2017YFA0206304, the National Natural Science Foundation of China under grant no. U1732159, 11274003, 91421109, 11522432, 61427812 and 11574288, the Priority Academic Program Development of Jiangsu Higher Education Institutions, and Collaborative Innovation Center of Solid-State Lighting and Energy-Saving Electronics.

# Supporting Information

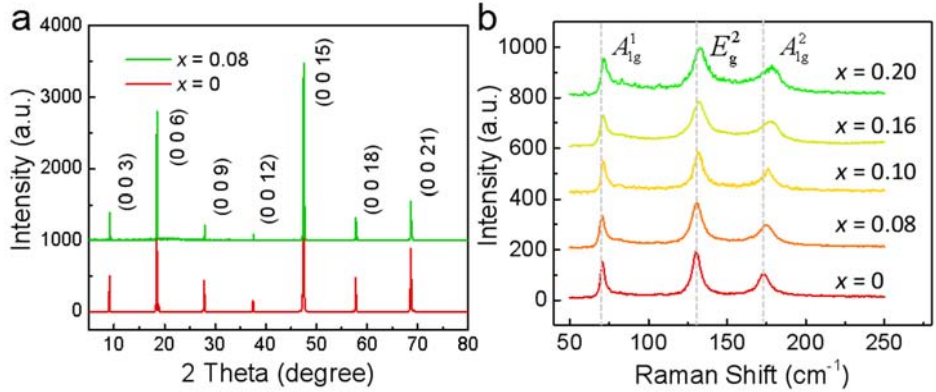

**Figure S1.** The structure characterization of $(Bi_{1-x}In_x)_2Se_3$. (a) X-ray diffraction patterns of bulk $(Bi_{1-x}In_x)_2Se_3$. $(Bi_{1-x}In_x)_2Se_3$ ($x = 0.08$) has the similar diffraction pattern with that of the pristine $Bi_2Se_3$. Only sharp (00$l$) reflections are observed, confirming the excellent crystalline quality of the bulk single crystals. After the doping of In into $Bi_2Se_3$, the crystal structure is maintained a rhombohedral phase. (b) Raman spectra of $(Bi_{1-x}In_x)_2Se_3$ thin flakes. At $x = 0$, Raman spectrum shows three phonon peaks at 71, 130, and 173 cm$^{-1}$, corresponding to the $A_{1g}^1$, $E_g^2$, and $A_{1g}^2$ vibrational modes, respectively. Raman spectra of all the thin flakes are similar to each other, indicating that the local crystalline order is maintained in spite of the In incorporation.

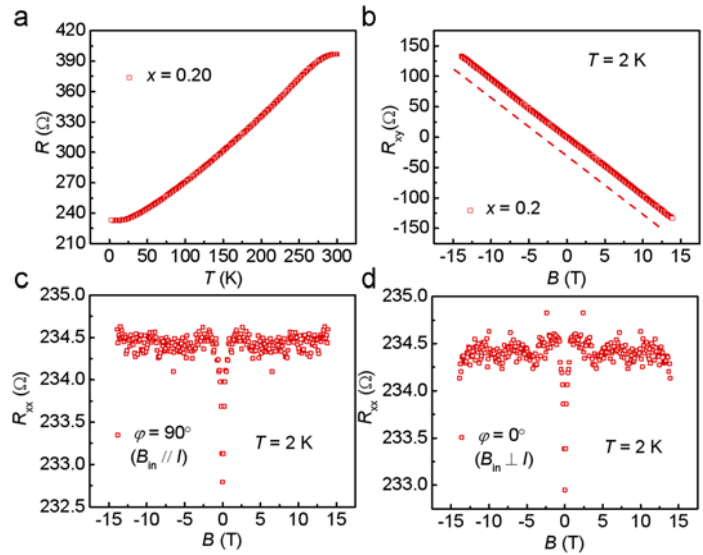

**Figure S2.** The in-plane positive MR of $(Bi_{1-x}In_x)_2Se_3$ thin flakes ($x = 0.20$) at 2 K. (a) Temperature-dependent resistance curve. (b) The linear Hall resistance in the out-of-plane magnetic field ($\theta = 0°$) at 2 K, suggesting the disappearance of the surface channel. (c and d) The positive MR of $(Bi_{1-x}In_x)_2Se_3$ thin flakes ($x = 0.20$) at 2 K, respectively, both in the in-plane magnetic field ($\varphi = 0°$ and 90°).

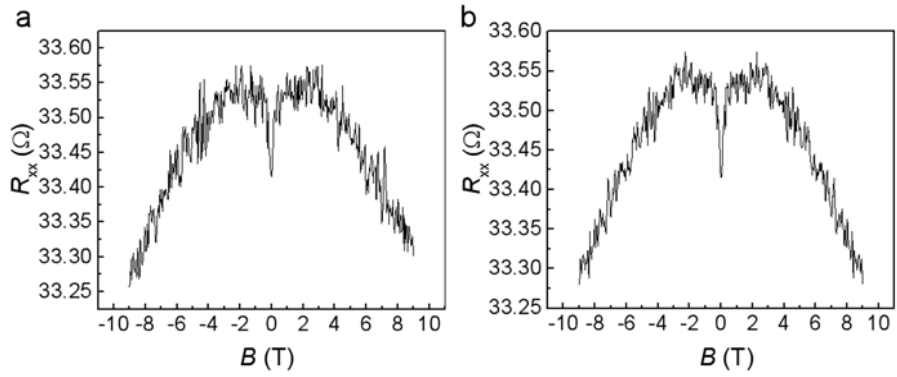

**Figure S3.** Negative MR in Nanodevice 3 measured by 2-terminal geometry. (a) The raw magnetotransport data up to 9 T, which shows the negative MR. The contact resistance is very small. (b) The magnetotransport data after symmetrization.

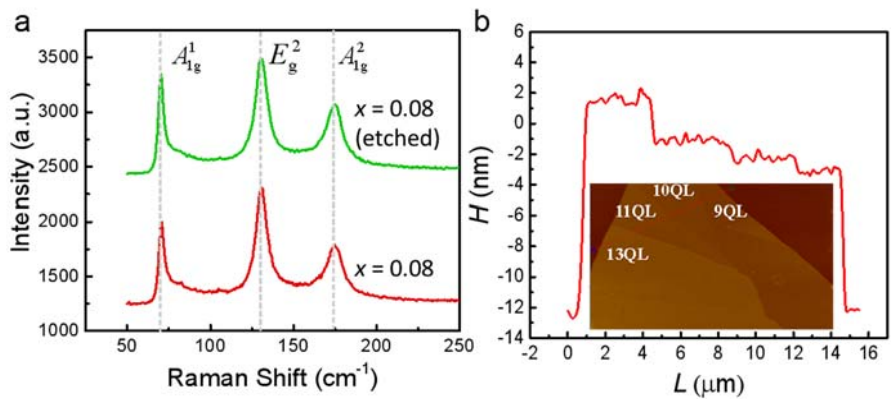

**Figure S4.** Raman spectra of $(Bi_{1-x}In_x)_2Se_3$ thin flakes ($x = 0.08$) with/without plasma exposure. (a) Raman spectra of the thin flakes are similar to each other, indicating that the local crystalline order is maintained in spite of the argon plasma etching. (b) An atomic force microscopy (AFM) line profile across the terraces that is marked with a line in the AFM image (inset). Distinct boundaries among the different thicknesses of ~9-13 quintuple layer (QL), can be clearly seen in the AFM image across the whole thin flake. Plasma exposure is found to reduce the thickness in a nonuniform way according to the most etched samples we studied. The surface thus becomes inhomogeneous after plasma exposure, reaching a roughness of ~1 nm.

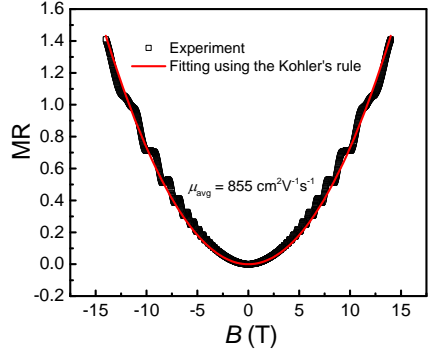

**Figure S5.** Nonsaturating parabolic magnetoresistance (MR) of the bulk single crystal ($x = 0.08$) at 2 K. The MR is nonsaturating and increased in proportion to the square of the magnetic field. An average mobility of 855 cm$^2$V$^{-1}$s$^{-1}$ is obtained in the out-of-plane magnetic field via fitting using the Kohler's rule (see the red line).

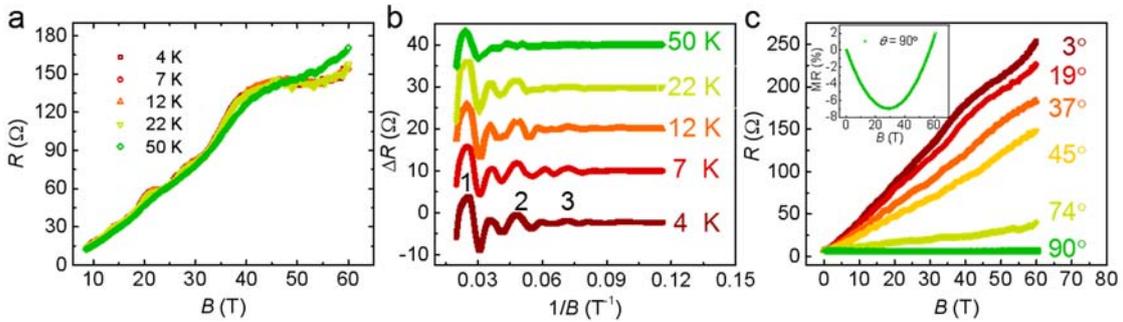

**Figure S6.** Magnetotransport results of composition at $x = 0.08$ in the high-pulsed magnetic field. (a) Temperature-dependent Shubnikov-de Haas (SdH) oscillations up to $B = 60$ T. (b) SdH oscillations of surface states after subtracting a smooth background of (a). It is seen that a quantum limit is achieved at $B = \sim 50$ T. The results indicate that the electron transport remains in the semi-classical limit at $B \leq 14$ T shown in the main text. (c) The $\theta$ angle-dependent MR up to $B = 60$ T at 4 K. The inset shows the enlarged negative MR at $\theta = 90°$ ($B // I_{DS}$), which is maintained up to $B = \sim 25$ T and disappears afterwards.

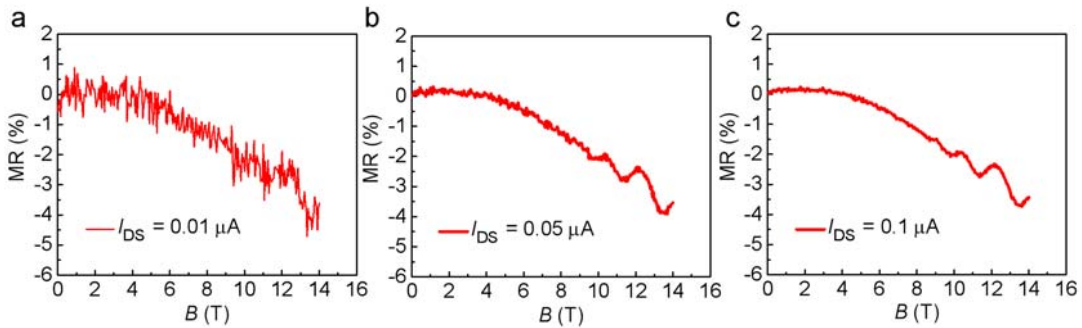

**Figure S7.** Current-density dependence of the in-plane negative MR at 2 K. It is seen that the variation of current density does not change the amplitude of the negative MR. It agrees with our proposed model (Figure 5) since the net spin polarization ($S_{net}$) is unrelated to the current density.